\documentclass[letterpaper]{article} 
\usepackage{aaai24}  
\usepackage{times}  
\usepackage{helvet}  
\usepackage{courier}  
\usepackage[hyphens]{url}  
\usepackage{graphicx} 
\usepackage{natbib}  
\usepackage{caption} 
\usepackage{algorithm}
\usepackage{algorithmic}
\usepackage{multirow}
\usepackage{booktabs}
\usepackage{amsmath}
\usepackage{amssymb}

\usepackage{newfloat}
\usepackage{listings}

\DeclareCaptionStyle{ruled}{labelfont=normalfont,labelsep=colon,strut=off} 
\floatstyle{ruled}
\newfloat{listing}{tb}{lst}{}
\floatname{listing}{Listing}
\title{StyleSinger: Style Transfer for Out-of-Domain Singing Voice Synthesis}
\author {
    Yu Zhang\textsuperscript{\rm 1},
    Rongjie Huang\textsuperscript{\rm 1},
    Ruiqi Li\textsuperscript{\rm 1},
    JinZheng He\textsuperscript{\rm 1},
    Yan Xia\textsuperscript{\rm 1},
    Feiyang Chen\textsuperscript{\rm 2},
    Xinyu Duan\textsuperscript{\rm 2},
    Baoxing Huai\textsuperscript{\rm 2},
    Zhou Zhao\textsuperscript{\rm 1}\thanks{Corresponding author}
}
\affiliations {
    \textsuperscript{\rm 1}Zhejiang University\\
    \textsuperscript{\rm 2}Huawei Cloud\\
    
    \{yuzhang34,rongjiehuang,ruiqili,jinzhenghe,zhaozhou\}@zju.edu.cn, xiayan.zju@gmail.com    
    \{chenfeiyang2,duanxinyu,huaibaoxing\}@huawei.com
}

\begin{document}

\maketitle

\begin{abstract}

Style transfer for out-of-domain (OOD) singing voice synthesis (SVS) focuses on generating high-quality singing voices with unseen styles (such as timbre, emotion, pronunciation, and articulation skills) derived from reference singing voice samples.
However, the endeavor to model the intricate nuances of singing voice styles is an arduous task, as singing voices possess a remarkable degree of expressiveness. 
Moreover, existing SVS methods encounter a decline in the quality of synthesized singing voices in OOD scenarios, as they rest upon the assumption that the target vocal attributes are discernible during the training phase.
To overcome these challenges, we propose StyleSinger, the first singing voice synthesis model for zero-shot style transfer of out-of-domain reference singing voice samples. 
StyleSinger incorporates two critical approaches for enhanced effectiveness: 
1) the Residual Style Adaptor (RSA) which employs a residual quantization module to capture diverse style characteristics in singing voices, and
2) the Uncertainty Modeling Layer Normalization (UMLN) to perturb the style attributes within the content representation during the training phase and thus improve the model generalization. 
Our extensive evaluations in zero-shot style transfer undeniably establish that StyleSinger outperforms baseline models in both audio quality and similarity to the reference singing voice samples.
Access to singing voice samples can be found at \url{https://aaronz345.github.io/StyleSingerDemo/}.
Code can be found at \url{https://github.com/AaronZ345/StyleSinger}.

\end{abstract}

\section{Introduction}

\begin{figure}[t]
\centering
\includegraphics[width=0.45\textwidth]{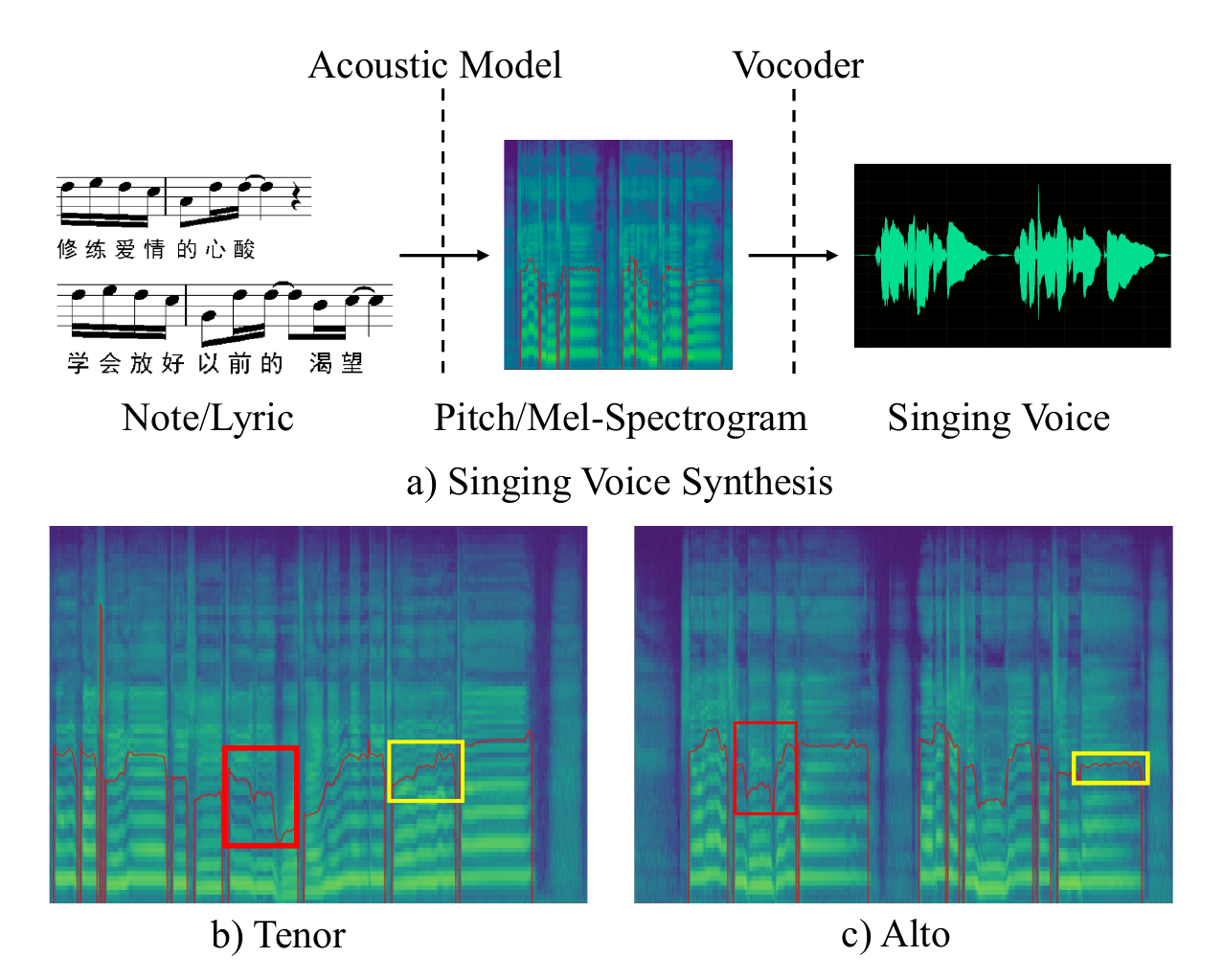}
\caption{Figure (a) shows the singing voice synthesis overall pipeline.
SVS systems use an acoustic model to transform musical notations and lyrics into intermediate features (like pitch and mel-spectrograms), and then a vocoder synthesizes the target singing voices.
In this paper, our method mainly focuses on the acoustic model.
Figures (b) and (c) depict the constituent elements of singing voice styles, namely pronunciation and articulation skills. 
Red boxed showcases pitch transitions and yellow boxes highlight the vibrato skill.}
\label{fig: style}
\end{figure}

Singing voice synthesis (SVS) is dedicated to generating high-quality singing voices through the utilization of lyrics and musical notations. 
This domain has witnessed significant advancements, finding crucial applications in both the realm of professional music composition and entertainment short videos. 
Currently, numerous outstanding SVS techniques demonstrate remarkable efficacy in synthesizing exceptional results \citep{zhang2022visinger, choi2022melody, kim2023muse, huang2022singgan, huang2021multi, he2023rmssinger}.

With the rapid development of SVS methods, there is a growing demand for out-of-domain (OOD) style transfer in singing voices, which seeks to generate high-quality singing voices with unseen styles derived from reference singing voice samples.
To be more specific, styles of singing voices primarily include {\bf timbre, emotion, pronunciation, and articulation skills}. 
Timbre represents the fundamental and distinctive quality of a singer's voice, while emotion captures the expressive and emotional delivery conveyed during a performance. 
As shown in Figure \ref{fig: style}(b) and (c), pronunciation and articulation skills involve various techniques such as vibrato, pitch transitions, and enunciation skills. 
However, current SVS systems lack the necessary techniques to effectively model the intricate styles of singing voices.  
Consequently, existing SVS methods encounter a decline in the quality of synthesized samples in OOD scenarios, as they rest upon the assumption that the target vocal attributes are discernible during the training phase.

In essence, the challenges of the style transfer for OOD SVS can be summarized as follows: 
1) Modeling the intricate nuances of singing voice styles is an arduous task, as singing voices possess a remarkable degree of expressiveness. 
Some methods for style modeling utilize a speaker encoder \citep{kumar2021normalization}. 
Moreover, other approaches model styles from multiple perspectives \citep{casanova2022yourtts}. 
However, these methods only consider limited aspects of speech styles and do not model the detailed styles of singing voices, such as pronunciation and articulation skills.
2) Disparities between the styles of OOD reference samples and the training data often lead to a deterioration in the quality of the synthesized singing voices. 
Many methods for model generalization rely on extensive data \citep{jia2018transfer}, which will be costly for singing voices. 
Alternatively, some methods employ a style adaptor for unseen styles \citep{min2021meta}, but they often require direct access to the target voice for model adaptation, which is not always feasible.

To tackle these challenges, we propose StyleSinger, the first singing voice synthesis (SVS) model for zero-shot style transfer of out-of-domain (OOD) reference samples. 
To capture the diverse style information in singing voices, we introduce the Residual Style Adaptor (RSA).
The RSA employs a residual quantization module to capture detailed style characteristics (e.g., pronunciation and articulation skills) in reference samples.
To improve the model generalization, we propose the Uncertainty Modeling Layer Normalization (UMLN).
The UMLN perturbs the style attributes within the content representation during the training phase, so the model performs better when faced with unseen reference styles during testing.
Our comprehensive evaluations in zero-shot style transfer establish that StyleSinger surpasses the baseline models in singing quality and similarity to the reference style.
The main contributions of this work are:

\begin{itemize}
    \item We present StyleSinger, the first singing voice synthesis model for zero-shot style transfer of out-of-domain reference samples. 
    StyleSinger excels in generating exceptional singing voices with unseen styles derived from reference singing voice samples.
    \item We propose the Residual Style Adaptor (RSA), which uses a residual quantization model to meticulously capture diverse style characteristics in reference samples.
    \item We introduce the Uncertainty Modeling Layer Normalization (UMLN) to perturb the style information in the content representation during the training phase, and thus enhance the model generalization of StyleSinger.
    \item Extensive experiments in zero-shot style transfer show that StyleSinger exhibits superior audio quality and similarity compared with baseline models.
\end{itemize}

\section{Related Works}

\subsection{Singing Voice Synthesis}

Singing voice synthesis (SVS) aims to generate singing voices of exceptional quality based on provided musical scores and lyrics. 
DiffSinger \citep{liu2022diffsinger} introduces a diffusion decoder \citep{ho2020denoising} to generate high-fidelity mel-spectrograms. 
In the multi-singer scenarios, MuSE-SVS \citep{kim2023muse} presents a multi-singer emotional singing voice synthesizer. 
M4Singer \citep{zhang2022m4singer} releases a multi-style, multi-singer Chinese song corpus with meticulously annotated fine-grained music scores.
Wesinger \citep{zhang2022wesinger} proposes a Transformer-alike acoustic model.
Recently, RMSSinger \citep{he2023rmssinger} proposes a method based on realistic music scores, utilizing a diffusion pitch prediction model to forecast F0 and UV. 
However, these SVS methods encounter challenges in maintaining synthesis quality when dealing with out-of-domain singers and styles, as well as in accurately modeling the intricate nuances of singing voice styles. 
In this paper, our approach successfully tackles these difficulties.

\subsection{Style Modeling}

The field of audio research has dedicated significant efforts to the exploration of style modeling. 
Attentron \citep{choi2020attentron} introduces an attention mechanism to extract styles from reference samples. 
\citet{cooper2020zero} proposes a speaker embedding method to model the reference samples.
ZSM-SS \citep{kumar2021normalization} proposes a Transformer-based architecture with an external speaker encoder using wav2vec 2.0 \citep{baevski2020wav2vec}. 
Moreover, numerous methods focus on modeling multi-level audio styles apart from speaker embedding.
\citep{li2021towards} incorporates global utterance-level and local phoneme-level style features in target speech. 
SC-GlowTTS \citep{casanova2021sc} presents a speaker-conditional architecture utilizing flow-based models.
Meta-StyleSpeech \citep{min2021meta} employs a speech encoding network for synthesizing multi-speaker TTS. 
Styler \citep{lee2021styler} disentangles style factors with equal supervision levels. 
Generspeech \citep{huang2022generspeech} incorporates both global and local style adaptors to capture styles. 
However, these approaches focus on limited aspects of speech styles and fail to capture the pronunciation and articulation skills of singing voice styles.

\subsection{Model Generalization}

Enabling the model to effectively capture the essence of unfamiliar out-of-domain test data presents a formidable challenge that SVS models must confront. 
Prominent methodologies \citep{jia2018transfer, paul2020speaker} leverage extensive data to achieve generalization. 
When it comes to singing voice data, acquiring a substantial amount of annotated data proves to be both costly and arduous. 
\citet{min2021meta, huang2022meta} employ meta-learning as the style adaptor for unseen speakers not encountered during the training phase. 
Such style adaptation methods require accessibility to the target voice, which is not always feasible.
In contrast, \citet{casanova2022yourtts} have devised an architecture that builds upon VITS, yielding exceptional zero-shot results. 
In the image domain, certain approaches focus on manipulating feature statistics to improve model generalization. 
MixStyle \citep{zhou2021domain} utilizes linear interpolation on feature statistics and shuffles the input samples to generate synthesized samples. 
Similarly, pAdaIn \citep{nuriel2021permuted} applies a random permutation to swap sample statistics.
Nevertheless, all of these approaches primarily concentrate on the domains of speech or image, whereas our focus is on the realm of singing voices.

\begin{figure*}[t]
\centering
\includegraphics[width=0.9\textwidth]{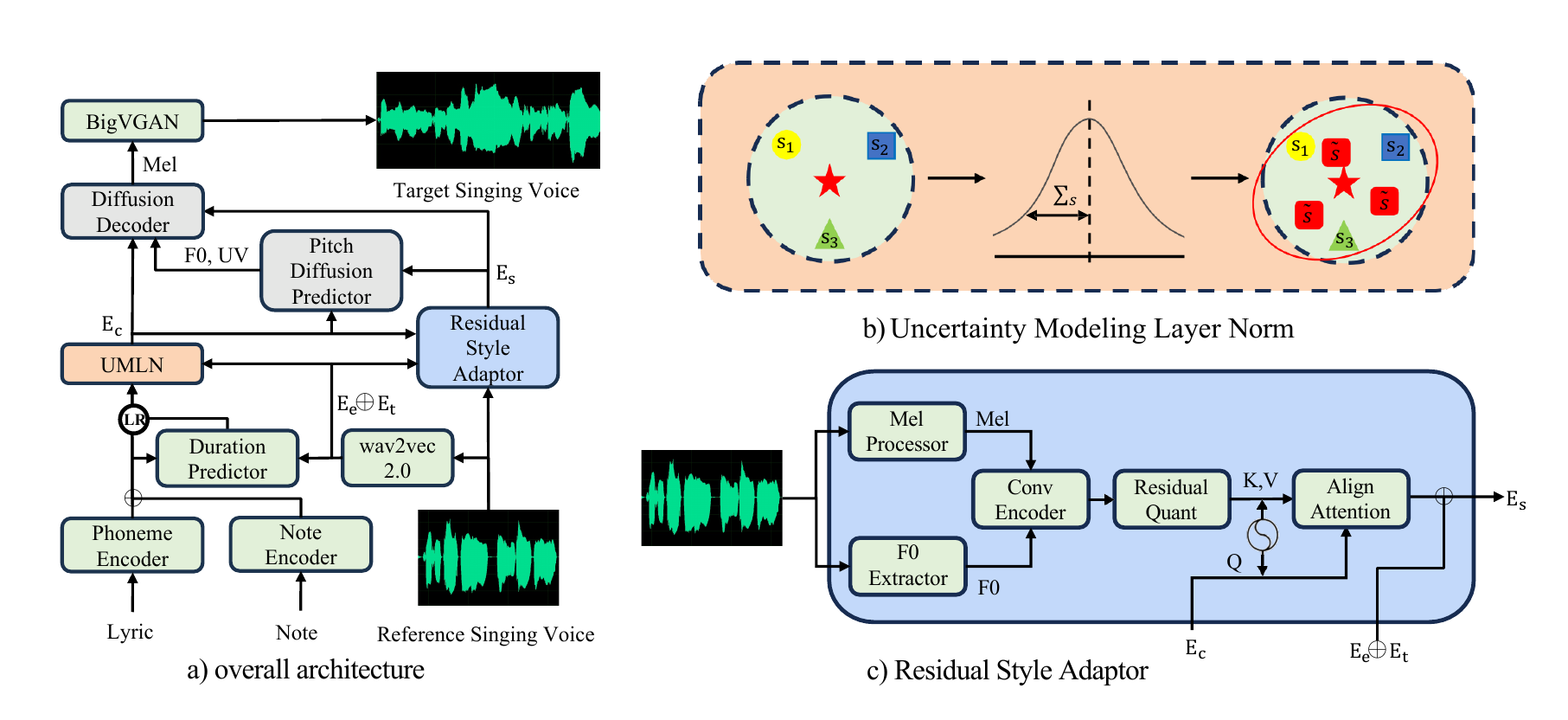}
\caption{The architecture of StyleSinger. 
In Figure (a), UMLN is the Uncertainty Modeling Layer Normalization. 
LR means length regulator.
$E_t$ and $E_e$ represent the embedding of timbre and emotion respectively, while $E_c$ and $E_s$ denote the style-agnostic representation and style-specific representation. 
In Figure (b), $s$ and $\tilde{s}$ are the input and output style information. 
In Figure (c), mel-spectrograms and f0 are extracted from the reference singing voice.}
\label{fig: arch}
\end{figure*}

\section{StyleSinger}

In this section, we first define the task of style transfer for out-of-domain singing voice synthesis.
Then we overview the proposed StyleSinger. 
After that, we introduce several critical components including the Uncertainty Modeling Layer Normalization (UMLN), the Residual Style Adaptor (RSA), and architectural details. 
Finally, we elaborate on the pre-training, training, and inference pipeline of StyleSinger.

\subsection{Problem Formulation}

Given target lyrics and notes, the objective of style transfer for out-of-domain (OOD) singing voice synthesis (SVS) is to generate high-quality target singing voices with unseen styles (such as timbre, emotion, pronunciation, and articulation skills) extracted from reference singing voice samples.

\subsection{Overview}

The architecture of StyleSinger is illustrated in Figure \ref{fig: arch}(a).
Lyrics are encoded through the phoneme encoder, while the note encoder captures musical notes. 
To extract timbre and emotion embedding from the reference singing voice, we utilize a pre-trained wave2vec 2.0 \citep{baevski2020wav2vec}. 
Then we split our model into style-agnostic and style-specific parts to achieve better generalization \citep{li2017deeper, li2019feature}.
After predicting the duration, we utilize the Uncertainty Modeling Layer Normalization (UMLN) to perturb the style information in the content representation. 
This approach enhances the model generalization of StyleSinger and acquires the style-agnostic representation.
The reference singing voice is then processed by the Residual Style Adaptor (RSA), which employs a residual quantization module to capture detailed style information (such as pronunciation and articulation skills) and thus gets the style-specific representation. 
Subsequently, the pitch diffusion predictor gets both style-agnostic and style-specific representations as inputs to generate F0 and UV. 
The diffusion decoder then generates mel-spectrograms. 
Finally, the target singing voice is generated by BigVGan \citep{lee2022bigvgan}.

\subsection{Uncertainty Modeling Layer Normalization}
\label{sec: umln}

In general, the style vector is commonly incorporated into the generator by concatenating it with the encoder output. 
However, this approach can lead to a decline in model performance when encountering OOD scenarios. 
To address this issue, \citet{chen2021adaspeech} introduces conditional layer normalization for style adaptation, allowing for scaling and shifting of the normalized input features based on the style embedding. 
In this work, we propose the Uncertainty Modeling Layer Normalization (UMLN), which enhances the generalization performance of StyleSinger by incorporating regularization techniques that introduce perturbations to the style information in training samples.

To be more detailed, we can compute the mean $\mu$ and variance $\delta$ with a hidden vector $x$. 
Additionally, given the style vector $s$, we utilize two simple linear layers to convert the vector into the bias vector $\beta(s)$ and scale vector $\gamma(s)$. 
To perturb style information, we utilize a Gaussian distribution to model the uncertainty scope of style embedding. 
By sampling from the uncertainty scope, we can simulate a wide range of diverse unseen style information and effectively prevent the model from generating style-consistent representations.
Notably, several studies \citep{shen2021closed,wang2019implicit} have showcased that the variances observed within features bear implicit semantic connotations. 
To capture the uncertainties inherent in style embedding, we calculate the variances of the scale and bias vectors:
\begin{equation}
\begin{aligned}
&\Sigma^2_\gamma(s)=\frac{1}{B}\sum^B_{b=1}(\gamma(s)-\mathbb{E}_b[\gamma(s)])^2,\\
&\Sigma^2_\beta(s)=\frac{1}{B}\sum^B_{b=1}(\beta(s)-\mathbb{E}_b[\beta(s)])^2,
\end{aligned}
\end{equation}
where $\Sigma_\gamma$ and $\Sigma_\beta$ represent the uncertainty estimation of style embedding $s$. 
The magnitudes of uncertainty estimation provide the potential transformations that may transpire within the style embedding. 

As shown in Figure \ref{fig: arch}(b), we employ random sampling to perturb the style information in training samples and foster the cultivation of a style-agnostic representation. 
Drawing inspiration from the previous work \citep{li2022uncertainty}, we update the scale and bias vectors:
\begin{equation}
\begin{aligned}
&\gamma_{um}(s)=\gamma(s)+\epsilon_{\gamma}\Sigma^2_\gamma(s),\\
&\beta_{um}(s)=\beta(s)+\epsilon_{\beta}\Sigma^2_\beta(s),
\end{aligned}
\end{equation}
where $\epsilon_{\gamma}$ and $\epsilon_{\beta}$ are drawn from the standard Gaussian distribution $\mathcal{N}(0,1)$. 
Upon updating the scale and bias vectors, the style-agnostic hidden representation becomes:
\begin{equation}
\begin{aligned}
&UMLN(x,s)=\gamma_{um}(s) \frac{x-\mu(x)}{\delta(x)}+\beta_{um}(s).\\
\end{aligned}
\end{equation}

Ultimately, the model assiduously refines the input features, so attains style-agnostic representation. 
To strike a delicate balance within this module, we introduce a hyper-parameter $p$, which denotes the probability of using UMLN during the training phase. 
For the pseudo-code of the algorithm, please refer to Algorithm \ref{alg: umln} provided in Appendix \ref{sec: appendix2}.

\subsection{Residual Style Adaptor}
\label{sec: rse}

To intricately model the singing voice styles, we firstly use a wav2vec 2.0 \citep{baevski2020wav2vec} to capture the timbre and emotion attributes. 
However, the complexity of styles in singing voices is remarkably high. 
So we propose the Residual Style Adaptor (RSA) to capture additional style information, like pronunciation and articulation skills. 

As illustrated in Figure \ref{fig: arch}(c), we extract and encode mel-spectrograms and f0 from the reference singing voice sample. 
In this process, we utilize parselmouth \citep{jadoul2018introducing} to extract f0 information. 
Subsequently, we employ a Residual Quantization (RQ) module \citep{lee2022autoregressive} to extract the detailed style features, which establishes an information bottleneck and effectively eliminates non-style information. 
RQ has typically been used in the image field. 
Due to the ability of RQ to extract multiple layers of information, it enables more comprehensive and detailed modeling of style information across various hierarchical levels. 
In more concrete terms, pronunciation and articulation skills encompass pitch transitions between musical notes and vibrato within a musical note, where the multi-level modeling capability of RQ is highly suitable.

To be specific, the conv encoder generates an output $E$. 
With a quantization depth of $N$, the RQ module represents $E$ as a sequence of $N$ ordered codes. 
Let $RQ_i(E)$ denote the process of representing $E$ as RQ code and extracting code embedding in $i$-th codebook. 
The representation of $E$ in the RQ module at depth $n \in [N]$ is denoted as $\hat{E}^n=\sum_{i=1}^n RQ_i(E)$.
To ensure that the input representation adheres to a discrete embedding, a commitment loss \citep{lee2022autoregressive} is employed:
\begin{equation}
\label{equ: Lc}
\begin{aligned}
&\mathcal{L}_{c} = \sum_{n=1}^N\left\|E-sg[\hat{E}^n]\right\|_{2}^{2},
\end{aligned}
\end{equation}
where the notation $sg$ represents the stop gradient operator. 
It is important to note that $\mathcal{L}_{c}$ is the cumulative sum of quantization errors across all $n$ iterations, rather than a single term. 
The objective is to ensure that $\hat{E}^n$ progressively reduces the quantization error of $E$ as the value of $n$ increases.

After generating the detailed style embedding in the RQ module, it becomes necessary to align the embedding with the content representation $E_c$. 
To achieve this, we introduce the Align Attention module, which incorporates the Scaled Dot-Product Attention mechanism \citep{vaswani2017attention}. 
Before feeding the detailed style embedding into the attention module, we include positional encoding embedding. 
In the attention module, $E_c$ serves as the query, while the detailed style embedding $E_d$ serves as both the key and value, and $d$ represents the dimensionality of the key and query:
\begin{equation}
\begin{aligned}
&Attention(Q, K, V) =Attention(E_c, E_d, E_d)\\
&=Softmax(\frac{E_c E_d^{T}}{\sqrt{d}}) E_d.\\
\end{aligned}
\end{equation}

In the end, we acquire the detailed style representation, which we integrate with the content representation, as well as the timbre and emotion embedding generated from wav2vec 2.0. 
This integration results in the attainment of the style-specific representation.

\subsection{Architectural Details}
\label{sec: ad}

The overall architecture is depicted in Figure \ref{fig: arch}(a), and we shall briefly introduce a few other crucial components apart from UMLN and RSA. 

\subsubsection{Encoder}

Our encoder consists of a note encoder and a phoneme encoder. 
To be more specific, the phoneme encoder adopts the architecture in FastSpeech2 \citep{ren2020fastspeech}, which accepts phonemes as input and yields phoneme features. 
Meanwhile, the note encoder handles musical scores. 
It takes note pitches, note types (including rest, slur, grace, etc.), and note duration as inputs, and results in note features.
We combine the note and phoneme features to form the content representation.
For more detailed information on the encoder, please refer to Appendix \ref{sec: appendix1en}.

\subsubsection{Pitch Diffusion Predictor} 

When confronted with ever-evolving and dynamic singing voices, simple pitch predictor approaches demonstrate limited effectiveness. 
To capture the diverse styles in singing voices, we introduce the pitch diffusion predictor. 
The pitch diffusion predictor consists of the style-specific pitch diffusion predictor and the style-agnostic pitch diffusion predictor, both of which adhere to the same architectural principles as the previous pitch diffusion model \citep{he2023rmssinger}. 
By combining the outputs of them, we obtain the final predictions for F0 and UV. 
The optimization of this module is achieved through the utilization of Gaussian diffusion loss and multinomial diffusion loss \citep{he2023rmssinger}.
For more details about the pitch diffusion predictor, please refer to Appendix \ref{sec: appendix1pitch}.

\subsubsection{Diffusion Decoder} 

The dynamic nature of singing voice poses a challenge for traditional mel decoders, as they can not effectively capture the nuances of mel-spectrograms in singing voices.
To tackle this challenge, we employ the diffusion decoder to generate mel-spectrograms.
In our approach, we adopt the structure of the teacher model from ProDiff \citep{huang2022prodiff}, a 4-step generator-based diffusion model.  
To train the diffusion decoder, we use both the Mean Absolute Error (MAE) loss and Structural Similarity Index (SSIM) loss \citep{wang2004image}.
For more details about the diffusion decoder, please refer to Appendix \ref{sec: appendix1de}.

\subsection{Pre-training, Training and Inference Procedures}

The final loss terms of StyleSinger consist of the following parts: 
1) Duration prediction loss $\mathcal{L}_{dur}$: MSE between the predicted and the GT phoneme-level duration in log scale; 
2) Pitch reconstruction loss $\mathcal{L}_{gdiff}, \mathcal{L}_{mdiff}$: Gaussian diffusion loss and multinomial diffusion loss between the GT and the pitch spectrogram predicted by the pitch diffusion predictor; 
3) RQ loss $\mathcal{L}_{c}$: the commitment loss for residual quantization layer; 
4) Mel reconstruction loss $\mathcal{L}_{mae}, \mathcal{L}_{ssim}$: MAE loss and SSIM loss of the diffusion decoder.

During the pre-training phase, we train the wav2vec 2.0 model to classify timbres and emotions by the AM soft-max loss. 
When training StyleSinger, the reference and target singing voices remain unchanged. 
During the inference stage, we input lyrics and notes of the target singing voice, and with unseen reference samples, we synthesize target singing voices with OOD reference styles.

\section{Experiments}

\subsection{Experimental Setup}

In this section, we first provide an overview of the datasets used in our study. 
Next, we present the implementation details of our StyleSinger. 
We then discuss the training and evaluation details for the task. 
Finally, we introduce the baseline models that we employed for comparison purposes.

\subsubsection{Dataset}

Currently, there are no publicly available SVS datasets with style information. 
In this endeavor, we collect and annotate a Chinese song corpus (including 12 singers and 20 hours) by recruiting professional singers in a professional recording studio. 
Additionally, to include more acoustic variation, we incorporate the M4Singer dataset \citep{zhang2022m4singer} (including 20 singers and 30 hours), which is used under license CC BY-NC-SA 4.0. 
Under the guidance of music experts, we manually annotate these datasets with vocal range and emotion labels, categorizing them into 8 classes: tenor happy, tenor sad, soprano happy, soprano sad, bass happy, bass sad, alto happy, and alto sad. 
Finally, we randomly designate 2 of these classes (tenor happy and alto sad) and 8 singers as unseen styles to evaluate StyleSinger in the OOD scenario, and then randomly select 20 sentences with unseen styles to construct the OOD testing set. 

\subsubsection{Implementation Details}

We utilize pypinyin to convert Chinese lyrics into phonemes. 
We extract mel-spectrograms from raw waveforms and set the sample rate to 48000Hz, the window size to 1024, the hop size to 256, and the number of mel bins to 80. 
The default size of the codebook in the RQ is set to 128, and the depth of the RQ is 4.
For more information, please refer to Appendix \ref{sec: appendix1arch}.

\subsubsection{Training Details}

We train our model for 20000 steps using 1 NVIDIA 2080Ti GPU. 
Adam optimizer is used with $\beta 1 = 0.9,\beta2 = 0.98$.  
It takes about 24 hours for training on 1 NVIDIA 2080Ti GPU.

\subsubsection{Evaluation Details}

In our experimental analysis, we employ both objective and subjective evaluation metrics to assess the synthesis quality and style similarity of the test set. 
For objective evaluation, we utilize the Speaker Cosine Similarity (Cos) to quantify the timbre resemblance between the synthesized and reference singing voices, and F0 Frame Error (FFE) to quantify the synthesis quality. 
Regarding subjective evaluation, we rely on the Mean Opinion Score (MOS) to gauge naturalness and employ the Similarity Mean Opinion Score (SMOS) \citep{min2021meta} to assess style similarity. 
Additionally, in the ablation study, we conduct Comparative Mean Opinion Score (CMOS) and Comparative Similarity Mean Opinion Score (CSMOS) evaluations. 
All these metrics are rated from 1 to 5 and reported with 95\% confidence intervals. 
Moreover, we employ an AXY test \citep{skerry2018towards} to evaluate the style transfer performance. 
We employ the BigVGAN \citep{lee2022bigvgan} for all experiments. 
For more detailed information on the evaluation process, please refer to Appendix \ref{sec: appendix3}.

\subsubsection{Baseline Models}

We conduct a comparative analysis of the quality and similarity of the singing voice samples generated by our esteemed StyleSinger system with other systems, encompassing the following: 
1) Reference: The original reference singing voice sample; 
2) Reference (vocoder): We transform the reference singing voice sample into mel-spectrograms and subsequently regenerate it into a singing voice using BigVGAN; 
3) Styler \citep{lee2021styler}: We incorporate a module for handling note embedding into Styler, enabling it to generate singing voice performances.
4) GenerSpeech \citep{huang2022generspeech}: within GenerSpeech, we add a note encoder enabling GenerSpeech to accomplish style transfer of singing voice performances; 
5) YourTTS \citep{casanova2022yourtts}: Incorporating the architecture of YourTTS, we likewise integrate a module for note embedding to process the singing voice data.
6) Multi-Style RMSSinger \citep{he2023rmssinger} (MS RMSSinger): we enrich the architecture of RMSSinger by integrating the timbre and emotion vectors extracted by wav2vec 2.0 into its backbone, allowing it to handle style transfer tasks.

\subsection{Main Results}

\begin{table*}[t]
\centering
\small
\vspace{2mm}
\scalebox{0.98}{
\begin{tabular}{l|cc|cc}
\toprule
\rule{0pt}{10pt} \bfseries{Method} & \bfseries{MOS} $\uparrow$ & \bfseries{SMOS} $\uparrow$ & \bfseries{Cos} $\uparrow$ & \bfseries{FFE} $\downarrow$ \\
\midrule  
Refernece & 4.53 $\pm$ 0.05 & - & - & -
\\
Reference (vocoder) & 4.13 $\pm$ 0.07 & 4.26 $\pm$ 0.09 & 0.97 & 0.05 \\
\midrule  
Styler \citep{lee2021styler} & 3.52 $\pm$ 0.08 & 3.79 $\pm$ 0.07 & 0.77 & 0.39
\\
GenerSpeech \citep{huang2022generspeech} & 3.59 $\pm$ 0.07 & 3.83 $\pm$ 0.08 & 0.83 & 0.36 
\\
YourTTS \citep{casanova2022yourtts} & 3.65 $\pm$ 0.09 & 3.85 $\pm$ 0.10 & 0.85 & 0.31 
\\
MS RMSSinger \citep{he2023rmssinger} & 3.84 $\pm$ 0.08 & 3.90 $\pm$ 0.11 & 0.88 & 0.29
\\
\midrule  
StyleSinger (ours) & \bf 3.90 $\pm$ 0.05 & \bf 4.03 $\pm$ 0.07 & \bf 0.90 & \bf 0.27
\\
\bottomrule      
\end{tabular}}
\caption{
The quality and style similarity of parallel style transfer when extended to out-of-domain test sets. 
For subjective measurement, we employ MOS and SMOS. 
In objective evaluation, we utilize Cos and FFE.
}
\label{tab: base}
\end{table*}

\begin{table*}[t]
\centering
\small
\vspace{2mm}
\scalebox{0.98}{
\begin{tabular}{l|c|ccc|c|ccc}
\toprule  
\multirow{3}{*}{\bfseries{Baseline}} & \multicolumn{4}{c|}{\textbf{Parallel}} & \multicolumn{4}{c}{\textbf{Non-Parallel}} \\ \cline{2-9}
\rule{0pt}{10pt}& \multirow{2}{*}{7-point score} & \multicolumn{3}{c|}{Perference (\%)} & \multirow{2}{*}{7-point score} & \multicolumn{3}{c}{Perference (\%)} \\ 
\rule{0pt}{10pt}      &    & X & Neutral & Y     &   & X   & Neutral  & Y  \\
\midrule
Styler &  1.31 $\pm$ 0.14  & 18\%  & 26\% &  56\%   & 1.40 $\pm$ 0.12   & 10\%  & 22\% &  68\% \\
GenerSpeech & 1.29 $\pm$ 0.11  & 28\%  & 20\% &  52\%   & 1.33 $\pm$ 0.09   & 12\%  & 24\% &  64\% \\
YourTTS &  1.28 $\pm$ 0.08  & 26\%  & 24\% &  50\%   & 1.30 $\pm$ 0.10   & 16\%  & 20\% &  64\% \\
MS RMSSinger &  1.14 $\pm$ 0.10  & 28\%  & 30\% &  42\%  & 1.26 $\pm$ 0.12 & 24\%  & 16\% &  60\%\\
\bottomrule  
\end{tabular}}
\caption{
The AXY preference test results for parallel and non-parallel style transfer are presented. 
From the testing sets, we have selected 20 samples for evaluation. 
Raters were requested to assign a 7-point score (ranging from -3 to 3) and select the samples that sounded closer to the target style. 
In this context, X represents a baseline model, while our StyleSinger is denoted as Y. 
A higher score indicates that Y is closer to the target style compared to X.}
\label{tab: axy}
\end{table*}

We randomly select singing voice samples from the OOD testing sets as references to assess the style transfer capabilities of StyleSinger and baseline models. 
Based on the content consistency between the reference and generated singing voices, we categorize the experiments into parallel and non-parallel style transfer \citep{skerry2018towards}.

\subsubsection{Parallel Style Transfer}

In the context of out-of-domain (OOD) scenarios, where the content of the reference voice remains unchanged, the primary outcomes are presented in Table \ref{tab: base}. 
Based on both objective and subjective evaluations, the following observations can be made: 
1) StyleSinger demonstrates exceptional audio quality, as evidenced by the highest Mean Opinion Score (MOS) among all models. 
This signifies the model's remarkable universality in handling out-of-domain (OOD) scenarios.
2) StyleSinger also excels in style similarity, as indicated by the highest Style Mean Opinion Score (SMOS). 
This showcases the model's exceptional ability to accurately model and capture the nuances of different singing styles.
3) As measured by the objective indicators Cos and FFE, StyleSinger consistently delivers the best results.
These findings collectively demonstrate the remarkable effectiveness of StyleSinger in OOD scenarios for singing voice synthesis and style transfer. 
This can be attributed to the exceptional generalization capability of UMLN, the adeptness of the RSA in modeling style representations, and the integration of the pitch diffusion predictor and the diffusion decoder, which imbue the generated OOD singing voices with enhanced details and vividness.
For more details, please refer to Appendix \ref{sec: appendix4par}.

\subsubsection{Non-Parallel Style Transfer}

\begin{figure*}[t]
\centering
\includegraphics[width=0.9\textwidth]{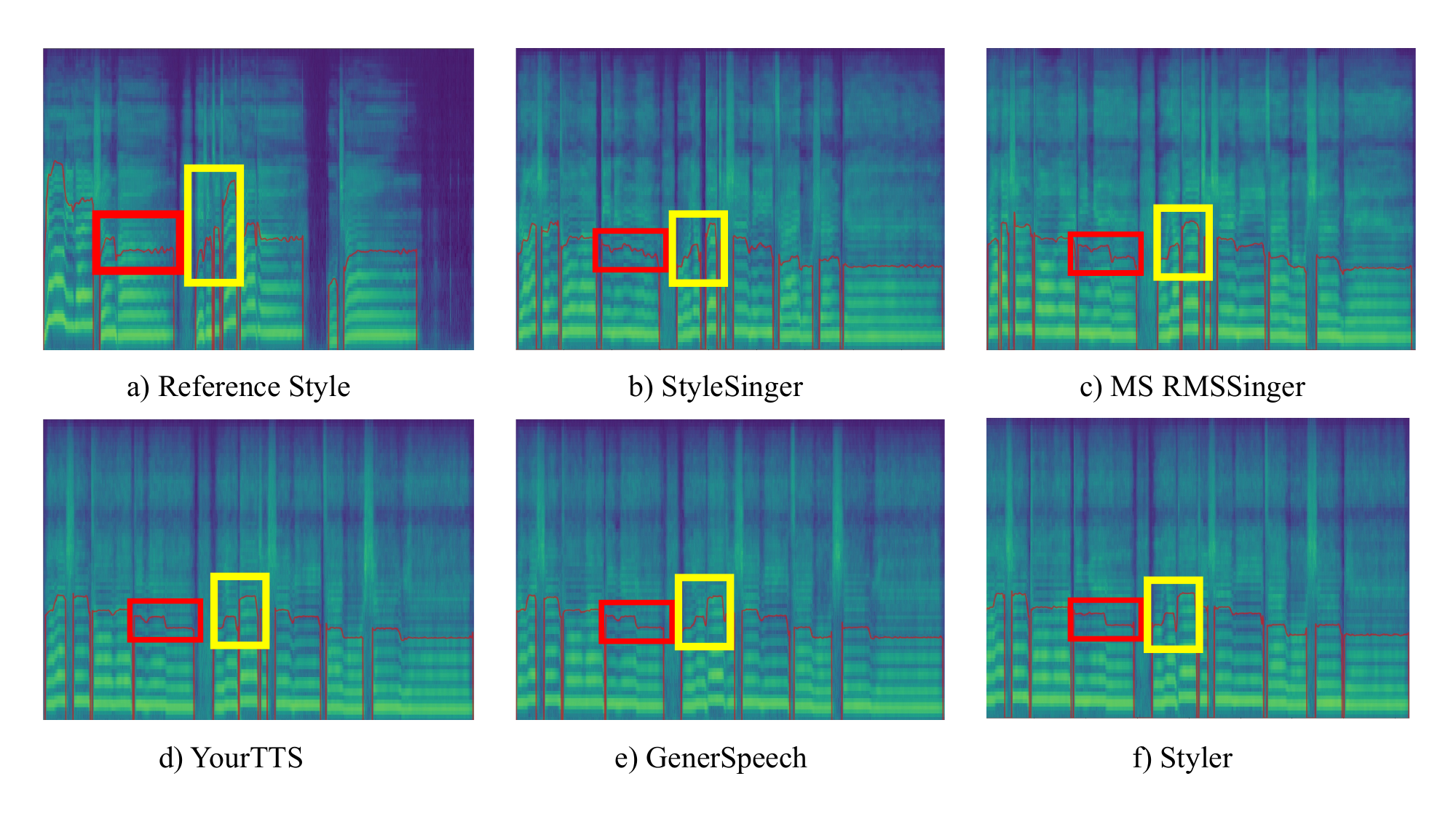}
\caption{The mel-spectrograms depicting the results of non-parallel style transfer. 
StyleSinger effectively captures the vibrato style indicated by red boxes, along with the pronunciation and articulation skills highlighted in yellow boxes.}
\label{fig: npa}
\end{figure*}

In out-of-domain (OOD) scenarios, we utilize unseen reference samples with target notes and lyrics to synthesize the target singing voice. 
To evaluate the performance, we conducted an AXY preference test by randomly selecting 20 unseen reference singing voice samples with target notes and lyrics. 
Then we compared the synthesis results of StyleSinger with baseline models. 
As shown in Table \ref{tab: axy}, the results demonstrate a clear preference for the synthesis generated by StyleSinger over the baselines. 
This affirms the effectiveness of our Residual Style Adaptor (RSA) and uncertainty modeling layer norm (UMLN) in achieving successful unseen style transfer.

We proceed to visualize mel-spectrograms and pitch contour in the context of non-parallel style transfer.
In Figure \ref{fig: npa}, it can be observed that:
1) StyleSinger excels at capturing the intricate nuances of the reference style. 
The pitch curve generated by StyleSinger displays a greater range of variations and finer details, closely resembling the reference style. 
To be more precise, StyleSinger effectively captures the vibrato style, as well as the nuances of pronunciation and articulation skills.
In contrast, the curves generated by other methods appear relatively flat, lacking distinctions in singing techniques.
2) StyleSinger excels in modeling mel-spectrograms of higher quality and intricate details. 
The mel-spectrograms generated by StyleSinger exhibit superior quality, showcasing rich details in frequency bins between adjacent harmonics and high-frequency components. 
In contrast, the mel-spectrograms generated by other methods for out-of-domain (OOD) samples demonstrate lower quality and a lack of intricate details.

When listening to the demo, it is evident that our model effectively captures the timbre, emotion, pitch transitions, vibrato, pronunciation, and articulation skills present in the reference singing voice samples.
Furthermore, it can be discerned that StyleSinger surpasses baseline models in synthesis quality and similarity to reference singing voice samples.

\subsection{Ablation Study}

\begin{table}[t]
\centering
\small
\vspace{2mm}
\begin{tabular}{l|c|c}
\toprule
\bfseries{Setting} & \bfseries{CMOS} & \bfseries{CSMOS} \\
\midrule
StyleSinger & 0.00  & 0.00 \\
\midrule
w/o UMLN  & -0.28  & -0.25 \\
w/o RSA  &  -0.12 & -0.29 \\
w/o Pitch  & -0.40  & -0.23 \\
w/o Decoder  & -0.52  & -0.19 \\
\bottomrule   
\end{tabular}
\caption{
Audio quality and similarity comparisons for ablation study with CMOS and CSMOS. 
UMLN and RSA are the Uncertainty Modeling Layer Normalization and the Residual Style Adaptor, while Pitch and Decoder mean the pitch diffusion predictor and the diffusion decoder.
}
\label{tab: abl}
\end{table}

As depicted in Table \ref{tab: abl}, we undertake ablation studies to showcase the efficacy of various designs incorporated within StyleSinger. 
We conduct CMOS (comparative mean opinion score) and CSMOS (comparative similarity mean opinion score) evaluations. 
1) When we eliminate the uncertainty modeling layer norm (UMLN), the quality and similarity decline, indicating the enhancement our method brings to model generalization performance. 
2) As the Residual Style Adaptor (RSA) is removed, the similarity significantly decreases, demonstrating the effectiveness of our method in modeling the intricate styles in singing voices.
3) Excluding the pitch diffusion predictor, we utilize the simple pitch predictor in FastSpeech2 \citep{ren2020fastspeech}, and the quality deteriorates further, highlighting the improvement our pitch predictor brings to f0 modeling.
4) Without the diffusion decoder, we employ a transformer decoder \citep{ren2020fastspeech} instead. The significant decline in audio quality highlights the crucial role of the diffusion decoder in generating high-quality mel-spectrograms.
For more detailed results of the ablation study, please refer to Appendix \ref{sec: appendix4abl}.

\section{Conclusion}

In this paper, we present a pioneering approach StyleSinger, the first singing voice synthesis model capable of achieving high-quality zero-shot style transfer for out-of-domain voices. 
We primarily enhance the model's performance through two key components: 
1) the Residual Style Adaptor (RSA) which employs a residual quantization module to capture diverse style characteristics in singing voices, and
2) the Uncertainty Modeling Layer Normalization (UMLN) to perturb the style attributes within the content representation during the training phase and thus improve the model generalization.
Our extensive evaluations in zero-shot style transfer undeniably establish that StyleSinger outperforms baseline models in both audio quality and similarity to the reference samples.
For future work, we aim to expand the capabilities of StylerSinger to encompass a broader range of scenarios, such as multilingual tasks. 
Additionally, it is a good idea to explore training models that integrate both speech and singing styles, which can generate singing voices with styles extracted from OOD reference speech.

\section*{Ethics Statement}

StyleSinger's ability to perform out-of-domain style transfer for singing voices raises concerns regarding potential unfair competition and the potential displacement of professionals within the music industry. 
Additionally, its application in the entertainment sector, including short videos, may give rise to copyright issues. 
Therefore, we will impose restrictions on our code and models to prevent unauthorized usage.

\section*{Acknowledgements}
This work was supported in part by the National Key R\&D Program of China under Grant No.2022ZD0162000, National Natural Science Foundation of China under Grant No.62222211, Grant No.61836002 and Grant No.62072397, and Yiwise.

\bibliography{aaai24}

\begin{thebibliography}{39}
\providecommand{\natexlab}[1]{#1}

\bibitem[{Baevski et~al.(2020)Baevski, Zhou, Mohamed, and Auli}]{baevski2020wav2vec}
Baevski, A.; Zhou, Y.; Mohamed, A.; and Auli, M. 2020.
\newblock wav2vec 2.0: A framework for self-supervised learning of speech representations.
\newblock \emph{Advances in neural information processing systems}, 33: 12449--12460.

\bibitem[{Casanova et~al.(2021)Casanova, Shulby, G{\"o}lge, M{\"u}ller, de~Oliveira, Junior, Soares, Aluisio, and Ponti}]{casanova2021sc}
Casanova, E.; Shulby, C.; G{\"o}lge, E.; M{\"u}ller, N.~M.; de~Oliveira, F.~S.; Junior, A.~C.; Soares, A. d.~S.; Aluisio, S.~M.; and Ponti, M.~A. 2021.
\newblock Sc-glowtts: an efficient zero-shot multi-speaker text-to-speech model.
\newblock \emph{arXiv preprint arXiv:2104.05557}.

\bibitem[{Casanova et~al.(2022)Casanova, Weber, Shulby, Junior, G{\"o}lge, and Ponti}]{casanova2022yourtts}
Casanova, E.; Weber, J.; Shulby, C.~D.; Junior, A.~C.; G{\"o}lge, E.; and Ponti, M.~A. 2022.
\newblock Yourtts: Towards zero-shot multi-speaker tts and zero-shot voice conversion for everyone.
\newblock In \emph{International Conference on Machine Learning}, 2709--2720. PMLR.

\bibitem[{Chen et~al.(2021)Chen, Tan, Li, Liu, Qin, Zhao, and Liu}]{chen2021adaspeech}
Chen, M.; Tan, X.; Li, B.; Liu, Y.; Qin, T.; Zhao, S.; and Liu, T.-Y. 2021.
\newblock Adaspeech: Adaptive text to speech for custom voice.
\newblock \emph{arXiv preprint arXiv:2103.00993}.

\bibitem[{Choi et~al.(2020)Choi, Han, Kim, and Ha}]{choi2020attentron}
Choi, S.; Han, S.; Kim, D.; and Ha, S. 2020.
\newblock Attentron: Few-shot text-to-speech utilizing attention-based variable-length embedding.
\newblock \emph{arXiv preprint arXiv:2005.08484}.

\bibitem[{Choi and Nam(2022)}]{choi2022melody}
Choi, S.; and Nam, J. 2022.
\newblock A melody-unsupervision model for singing voice synthesis.
\newblock In \emph{ICASSP 2022-2022 IEEE International Conference on Acoustics, Speech and Signal Processing (ICASSP)}, 7242--7246. IEEE.

\bibitem[{Cooper et~al.(2020)Cooper, Lai, Yasuda, Fang, Wang, Chen, and Yamagishi}]{cooper2020zero}
Cooper, E.; Lai, C.-I.; Yasuda, Y.; Fang, F.; Wang, X.; Chen, N.; and Yamagishi, J. 2020.
\newblock Zero-shot multi-speaker text-to-speech with state-of-the-art neural speaker embeddings.
\newblock In \emph{ICASSP 2020-2020 IEEE International Conference on Acoustics, Speech and Signal Processing (ICASSP)}, 6184--6188. IEEE.

\bibitem[{He et~al.(2023)He, Liu, Ye, Huang, Cui, Liu, and Zhao}]{he2023rmssinger}
He, J.; Liu, J.; Ye, Z.; Huang, R.; Cui, C.; Liu, H.; and Zhao, Z. 2023.
\newblock RMSSinger: Realistic-Music-Score based Singing Voice Synthesis.
\newblock \emph{arXiv preprint arXiv:2305.10686}.

\bibitem[{Ho, Jain, and Abbeel(2020)}]{ho2020denoising}
Ho, J.; Jain, A.; and Abbeel, P. 2020.
\newblock Denoising diffusion probabilistic models.
\newblock \emph{Advances in neural information processing systems}, 33: 6840--6851.

\bibitem[{Huang et~al.(2021)Huang, Chen, Ren, Liu, Cui, and Zhao}]{huang2021multi}
Huang, R.; Chen, F.; Ren, Y.; Liu, J.; Cui, C.; and Zhao, Z. 2021.
\newblock Multi-singer: Fast multi-singer singing voice vocoder with a large-scale corpus.
\newblock In \emph{Proceedings of the 29th ACM International Conference on Multimedia}, 3945--3954.

\bibitem[{Huang et~al.(2022{\natexlab{a}})Huang, Cui, Chen, Ren, Liu, Zhao, Huai, and Wang}]{huang2022singgan}
Huang, R.; Cui, C.; Chen, F.; Ren, Y.; Liu, J.; Zhao, Z.; Huai, B.; and Wang, Z. 2022{\natexlab{a}}.
\newblock Singgan: Generative adversarial network for high-fidelity singing voice generation.
\newblock In \emph{Proceedings of the 30th ACM International Conference on Multimedia}, 2525--2535.

\bibitem[{Huang et~al.(2022{\natexlab{b}})Huang, Ren, Liu, Cui, and Zhao}]{huang2022generspeech}
Huang, R.; Ren, Y.; Liu, J.; Cui, C.; and Zhao, Z. 2022{\natexlab{b}}.
\newblock Generspeech: Towards style transfer for generalizable out-of-domain text-to-speech synthesis.
\newblock \emph{arXiv preprint arXiv:2205.07211}.

\bibitem[{Huang et~al.(2022{\natexlab{c}})Huang, Zhao, Liu, Liu, Cui, and Ren}]{huang2022prodiff}
Huang, R.; Zhao, Z.; Liu, H.; Liu, J.; Cui, C.; and Ren, Y. 2022{\natexlab{c}}.
\newblock Prodiff: Progressive fast diffusion model for high-quality text-to-speech.
\newblock In \emph{Proceedings of the 30th ACM International Conference on Multimedia}, 2595--2605.

\bibitem[{Huang et~al.(2022{\natexlab{d}})Huang, Lin, Liu, Chen, and Lee}]{huang2022meta}
Huang, S.-F.; Lin, C.-J.; Liu, D.-R.; Chen, Y.-C.; and Lee, H.-y. 2022{\natexlab{d}}.
\newblock Meta-tts: Meta-learning for few-shot speaker adaptive text-to-speech.
\newblock \emph{IEEE/ACM Transactions on Audio, Speech, and Language Processing}, 30: 1558--1571.

\bibitem[{Jadoul, Thompson, and De~Boer(2018)}]{jadoul2018introducing}
Jadoul, Y.; Thompson, B.; and De~Boer, B. 2018.
\newblock Introducing parselmouth: A python interface to praat.
\newblock \emph{Journal of Phonetics}, 71: 1--15.

\bibitem[{Jia et~al.(2018)Jia, Zhang, Weiss, Wang, Shen, Ren, Nguyen, Pang, Lopez~Moreno, Wu et~al.}]{jia2018transfer}
Jia, Y.; Zhang, Y.; Weiss, R.; Wang, Q.; Shen, J.; Ren, F.; Nguyen, P.; Pang, R.; Lopez~Moreno, I.; Wu, Y.; et~al. 2018.
\newblock Transfer learning from speaker verification to multispeaker text-to-speech synthesis.
\newblock \emph{Advances in neural information processing systems}, 31.

\bibitem[{Kim et~al.(2023)Kim, Kim, Jun, and Kim}]{kim2023muse}
Kim, S.; Kim, Y.; Jun, J.; and Kim, I. 2023.
\newblock MuSE-SVS: Multi-Singer Emotional Singing Voice Synthesizer that Controls Emotional Intensity.
\newblock \emph{IEEE/ACM Transactions on Audio, Speech, and Language Processing}.

\bibitem[{Kumar et~al.(2021)Kumar, Goel, Narang, and Lall}]{kumar2021normalization}
Kumar, N.; Goel, S.; Narang, A.; and Lall, B. 2021.
\newblock Normalization Driven Zero-Shot Multi-Speaker Speech Synthesis.
\newblock In \emph{Interspeech}, 1354--1358.

\bibitem[{Lee et~al.(2022{\natexlab{a}})Lee, Kim, Kim, Cho, and Han}]{lee2022autoregressive}
Lee, D.; Kim, C.; Kim, S.; Cho, M.; and Han, W.-S. 2022{\natexlab{a}}.
\newblock Autoregressive image generation using residual quantization.
\newblock In \emph{Proceedings of the IEEE/CVF Conference on Computer Vision and Pattern Recognition}, 11523--11532.

\bibitem[{Lee, Park, and Kim(2021)}]{lee2021styler}
Lee, K.; Park, K.; and Kim, D. 2021.
\newblock Styler: Style factor modeling with rapidity and robustness via speech decomposition for expressive and controllable neural text to speech.
\newblock \emph{arXiv preprint arXiv:2103.09474}.

\bibitem[{Lee et~al.(2022{\natexlab{b}})Lee, Ping, Ginsburg, Catanzaro, and Yoon}]{lee2022bigvgan}
Lee, S.-g.; Ping, W.; Ginsburg, B.; Catanzaro, B.; and Yoon, S. 2022{\natexlab{b}}.
\newblock Bigvgan: A universal neural vocoder with large-scale training.
\newblock \emph{arXiv preprint arXiv:2206.04658}.

\bibitem[{Li et~al.(2017)Li, Yang, Song, and Hospedales}]{li2017deeper}
Li, D.; Yang, Y.; Song, Y.-Z.; and Hospedales, T.~M. 2017.
\newblock Deeper, broader and artier domain generalization.
\newblock In \emph{Proceedings of the IEEE international conference on computer vision}, 5542--5550.

\bibitem[{Li et~al.(2022)Li, Dai, Ge, Liu, Shan, and Duan}]{li2022uncertainty}
Li, X.; Dai, Y.; Ge, Y.; Liu, J.; Shan, Y.; and Duan, L.-Y. 2022.
\newblock Uncertainty modeling for out-of-distribution generalization.
\newblock \emph{arXiv preprint arXiv:2202.03958}.

\bibitem[{Li et~al.(2021)Li, Song, Li, Wu, Jia, and Meng}]{li2021towards}
Li, X.; Song, C.; Li, J.; Wu, Z.; Jia, J.; and Meng, H. 2021.
\newblock Towards multi-scale style control for expressive speech synthesis.
\newblock \emph{arXiv preprint arXiv:2104.03521}.

\bibitem[{Li et~al.(2019)Li, Yang, Zhou, and Hospedales}]{li2019feature}
Li, Y.; Yang, Y.; Zhou, W.; and Hospedales, T. 2019.
\newblock Feature-critic networks for heterogeneous domain generalization.
\newblock In \emph{International Conference on Machine Learning}, 3915--3924. PMLR.

\bibitem[{Liu et~al.(2022)Liu, Li, Ren, Chen, and Zhao}]{liu2022diffsinger}
Liu, J.; Li, C.; Ren, Y.; Chen, F.; and Zhao, Z. 2022.
\newblock Diffsinger: Singing voice synthesis via shallow diffusion mechanism.
\newblock In \emph{Proceedings of the AAAI conference on artificial intelligence}, volume~36, 11020--11028.

\bibitem[{Min et~al.(2021)Min, Lee, Yang, and Hwang}]{min2021meta}
Min, D.; Lee, D.~B.; Yang, E.; and Hwang, S.~J. 2021.
\newblock Meta-stylespeech: Multi-speaker adaptive text-to-speech generation.
\newblock In \emph{International Conference on Machine Learning}, 7748--7759. PMLR.

\bibitem[{Nuriel, Benaim, and Wolf(2021)}]{nuriel2021permuted}
Nuriel, O.; Benaim, S.; and Wolf, L. 2021.
\newblock Permuted adain: Reducing the bias towards global statistics in image classification.
\newblock In \emph{Proceedings of the IEEE/CVF Conference on Computer Vision and Pattern Recognition}, 9482--9491.

\bibitem[{Paul, Pantazis, and Stylianou(2020)}]{paul2020speaker}
Paul, D.; Pantazis, Y.; and Stylianou, Y. 2020.
\newblock Speaker conditional WaveRNN: Towards universal neural vocoder for unseen speaker and recording conditions.
\newblock \emph{arXiv preprint arXiv:2008.05289}.

\bibitem[{Ren et~al.(2020)Ren, Hu, Tan, Qin, Zhao, Zhao, and Liu}]{ren2020fastspeech}
Ren, Y.; Hu, C.; Tan, X.; Qin, T.; Zhao, S.; Zhao, Z.; and Liu, T.-Y. 2020.
\newblock Fastspeech 2: Fast and high-quality end-to-end text to speech.
\newblock \emph{arXiv preprint arXiv:2006.04558}.

\bibitem[{Shen and Zhou(2021)}]{shen2021closed}
Shen, Y.; and Zhou, B. 2021.
\newblock Closed-form factorization of latent semantics in gans.
\newblock In \emph{Proceedings of the IEEE/CVF conference on computer vision and pattern recognition}, 1532--1540.

\bibitem[{Skerry-Ryan et~al.(2018)Skerry-Ryan, Battenberg, Xiao, Wang, Stanton, Shor, Weiss, Clark, and Saurous}]{skerry2018towards}
Skerry-Ryan, R.; Battenberg, E.; Xiao, Y.; Wang, Y.; Stanton, D.; Shor, J.; Weiss, R.; Clark, R.; and Saurous, R.~A. 2018.
\newblock Towards end-to-end prosody transfer for expressive speech synthesis with tacotron.
\newblock In \emph{international conference on machine learning}, 4693--4702. PMLR.

\bibitem[{Vaswani et~al.(2017)Vaswani, Shazeer, Parmar, Uszkoreit, Jones, Gomez, Kaiser, and Polosukhin}]{vaswani2017attention}
Vaswani, A.; Shazeer, N.; Parmar, N.; Uszkoreit, J.; Jones, L.; Gomez, A.~N.; Kaiser, {\L}.; and Polosukhin, I. 2017.
\newblock Attention is all you need.
\newblock In \emph{Advances in neural information processing systems}, 5998--6008.

\bibitem[{Wang et~al.(2019)Wang, Pan, Song, Zhang, Huang, and Wu}]{wang2019implicit}
Wang, Y.; Pan, X.; Song, S.; Zhang, H.; Huang, G.; and Wu, C. 2019.
\newblock Implicit semantic data augmentation for deep networks.
\newblock \emph{Advances in Neural Information Processing Systems}, 32.

\bibitem[{Wang et~al.(2004)Wang, Bovik, Sheikh, and Simoncelli}]{wang2004image}
Wang, Z.; Bovik, A.~C.; Sheikh, H.~R.; and Simoncelli, E.~P. 2004.
\newblock Image quality assessment: from error visibility to structural similarity.
\newblock \emph{IEEE transactions on image processing}, 13(4): 600--612.

\bibitem[{Zhang et~al.(2022{\natexlab{a}})Zhang, Li, Wang, Deng, Liu, Ren, He, Huang, Zhu, Chen et~al.}]{zhang2022m4singer}
Zhang, L.; Li, R.; Wang, S.; Deng, L.; Liu, J.; Ren, Y.; He, J.; Huang, R.; Zhu, J.; Chen, X.; et~al. 2022{\natexlab{a}}.
\newblock M4singer: A multi-style, multi-singer and musical score provided mandarin singing corpus.
\newblock \emph{Advances in Neural Information Processing Systems}, 35: 6914--6926.

\bibitem[{Zhang et~al.(2022{\natexlab{b}})Zhang, Cong, Xue, Xie, Zhu, and Bi}]{zhang2022visinger}
Zhang, Y.; Cong, J.; Xue, H.; Xie, L.; Zhu, P.; and Bi, M. 2022{\natexlab{b}}.
\newblock Visinger: Variational inference with adversarial learning for end-to-end singing voice synthesis.
\newblock In \emph{ICASSP 2022-2022 IEEE International Conference on Acoustics, Speech and Signal Processing (ICASSP)}, 7237--7241. IEEE.

\bibitem[{Zhang et~al.(2022{\natexlab{c}})Zhang, Zheng, Li, and Lu}]{zhang2022wesinger}
Zhang, Z.; Zheng, Y.; Li, X.; and Lu, L. 2022{\natexlab{c}}.
\newblock Wesinger: Data-augmented singing voice synthesis with auxiliary losses.
\newblock \emph{arXiv preprint arXiv:2203.10750}.

\bibitem[{Zhou et~al.(2021)Zhou, Yang, Qiao, and Xiang}]{zhou2021domain}
Zhou, K.; Yang, Y.; Qiao, Y.; and Xiang, T. 2021.
\newblock Domain generalization with mixstyle.
\newblock \emph{arXiv preprint arXiv:2104.02008}.

\end{thebibliography}

\clearpage

\appendix

\section{Details of Models}
\label{sec: appendix1}

\begin{figure*}[t]
\centering
\includegraphics[width=0.9\textwidth]{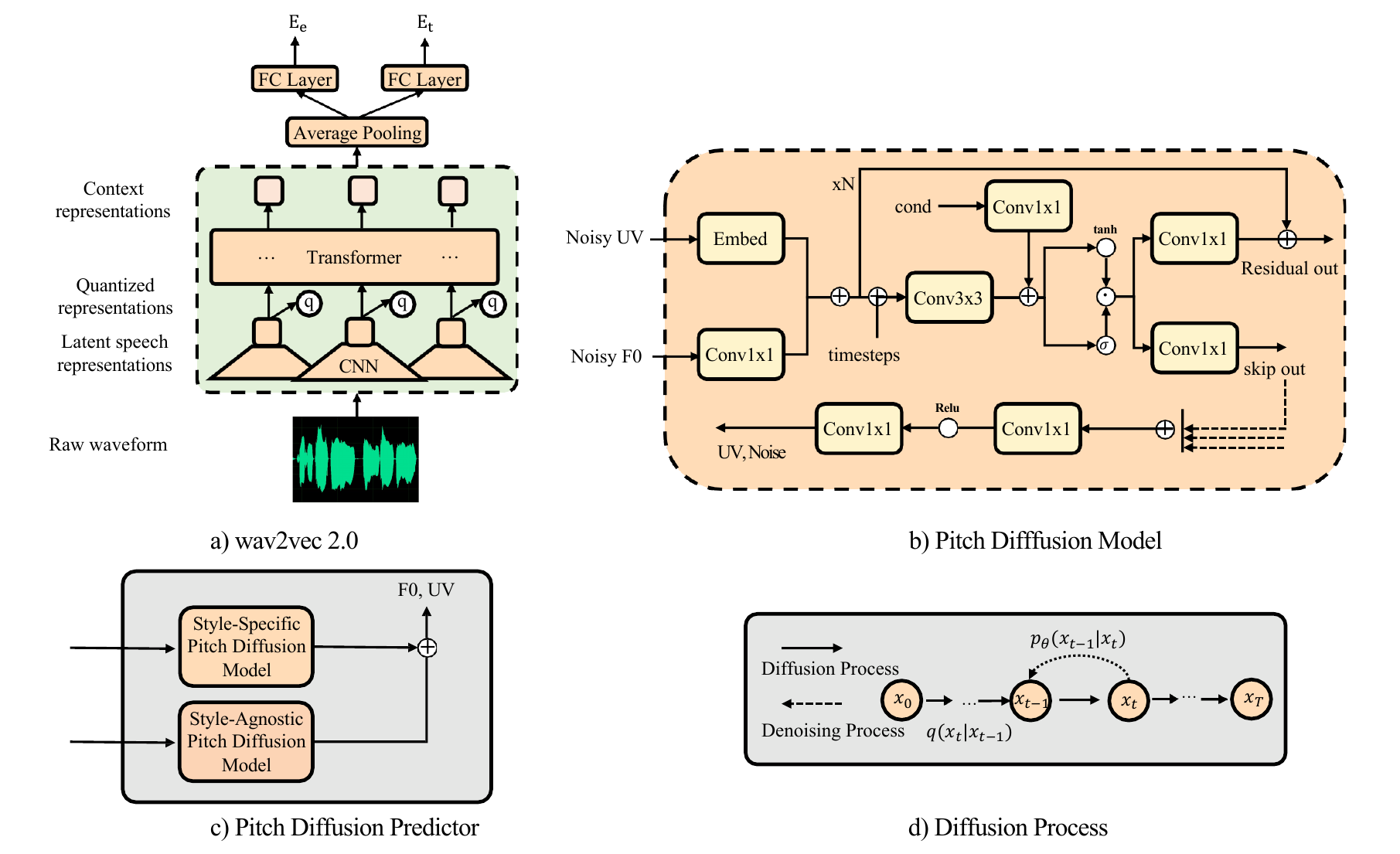}
\caption{Illustration of the downstream wav2vec 2.0, the pitch diffusion predictor and diffusion process. 
In Figure (a), we receive the awaiting classification waveform data as input, which, through the process of training, yields the esteemed timbre and emotion embedding. 
In Figure (b), the expanded note feature, expanded lyric feature, emotion, timbre, and detailed style embedding are summed and used as the condition. 
Figure (d) is a directed graph for diffusion models.}
\label{fig: dif}
\end{figure*}

\subsection{Architecture Details}
\label{sec: appendix1arch}

We list the architecture and hyperparameters in Table \ref{tab: arch}.

\begin{table}[t]
\centering
\small
\vspace{2mm}
\begin{tabular*}{\hsize}{l|c|c}
\toprule
\multicolumn{2}{c|}{\bfseries{Hyperparameter}}                            & \bfseries{StyleSinger}    \\
\midrule
\multirow{7}*{\shortstack{Phoneme\\Encoder}}            & Phoneme Embedding         & 256   \\
~                                                       & Encoder Layers            & 4     \\
~                                                       & Encoder Hidden            & 256   \\
~                                                       & Encoder Conv1D Kernel     & 9     \\
~                                                       & Encoder Conv1D Filter Size& 1024  \\
~                                                       & Encoder Attention Heads   & 2     \\
~                                                       & Encoder Dropout           & 0.1   \\
\midrule[0.2pt]
\multirow{3}*{\shortstack{Note\\Encoder}}               & Pitches Embedding         & 256   \\
~                                                       & Type Embedding            & 256   \\
~                                                       & Duration Hidden           & 256   \\
\midrule[0.2pt]
\multirow{1}*{\shortstack{UMLN}}                        & Probability of using UMLN & 0.5   \\
\midrule[0.2pt]
\multirow{4}*{\shortstack{Residual\\Style\\Adaptor}}    & Conv Encoder Layers       & 5     \\
~                                                       & RQ Codebook Size          & 128   \\
~                                                       & Depth of RQ               & 4   \\
~                                                       & Align Attention Layers    & 2     \\
\midrule[0.2pt]
\multirow{6}*{\shortstack{Pitch\\Diffusion\\Predictor}} & Conv Layers     & 12     \\
                                                        & Kernel Size               & 3      \\
                                                        & Residual Channel          & 192    \\
~                                                       & Hidden Channel            & 256    \\
~                                                       & Time Steps                & 100    \\
~                                                       & Max Linear $\beta$ Schedule& 0.06\\
\midrule[0.2pt]
\multirow{4}*{\shortstack{Diffusion\\Decoder}}          & Denoiser Layers           & 20     \\
~                                                       & Denoiser Hidden           & 256    \\
~                                                       & Time Steps                & 4      \\
~                                                       & Noise Schedule Type       & VPSDE  \\
\midrule[0.2pt]
\multicolumn{2}{c|}{Total Number of Parameters}                     & 42M    \\
\bottomrule
\end{tabular*}
\caption{
Hyper-parameters of StyleSinger modules.
}
\label{tab: arch}
\end{table}

\subsection{Encoder}
\label{sec: appendix1en}

Our encoder comprises a note encoder and a phoneme encoder. 
To elaborate, the phoneme encoder takes a sequence of phonemes as input. 
It passes through a phoneme embedding layer and four Feed-Forward Transformer (FFT) blocks, ultimately producing phoneme features. 
On the other hand, the note encoder handles musical score information. 
It takes note pitches, note types (including rest, slur, grace, etc.), and note duration as input. 
Note pitches, types, and duration undergo processing through two embedding layers and a linear projection layer respectively, resulting in the generation of note features.

\subsection{Wav2vec 2.0}
\label{sec: appendix1wav}

We employ wav2vec 2.0 for the task of classifying timbres and emotions. 
The architecture utilized in our approach is depicted in Figure \ref{fig: dif}(a).
The input waveform undergoes a series of transformations, including feature encoding through a CNN-based encoder, and network processing via a Transformer-based model with quantization modules, and culminates in a pooling layer, followed by two fully connected layers. 
These operations collectively yield the simultaneous generation of timbre and emotion embedding.

\subsection{Pitch Diffusion Predictor}
\label{sec: appendix1pitch}

As shown in Figure \ref{fig: dif}(c), the pitch diffusion predictor comprises the style-specific pitch diffusion predictor and the style-agnostic pitch diffusion predictor, both of which follow the same architectural principles as depicted in Figure \ref{fig: dif}(b). 
The diffusion process is illustrated in Figure \ref{fig: dif}(d).
These models incorporate both Gaussian diffusion and multinomial diffusion techniques to generate F0 and UV:
\begin{equation}
\begin{aligned}
&q(x_t|x_{t-1}) = \mathcal{N}(x_t;\sqrt{1-\beta_t}x_{t-1},\beta_t I),\\
&q(y_t|y_{t-1}) = \mathcal{C}(y_{t}|(1-\beta_t)y_{t-1}+\beta_t/K),
\end{aligned}
\end{equation}
where $\mathcal{C}$ represents a categorical distribution characterized by probability parameters, $x_t \sim \{0,1\}^K$, and $\beta_t$ denotes the probability of uniformly resampling a category. 
In the reverse process, the neural network is employed:
\begin{equation}
\begin{aligned}
&E_{x_0,\epsilon}[\frac{\beta_t^2}{2\sigma_t^2\alpha_t(1-\bar \alpha_t)}||\epsilon-\epsilon_\theta(x_t,t)||],\\
&q(y_{t-1}|y_t, y_0) = \mathcal{C}(y_{t-1}|\theta_{post}(y_t, y_0)),\\ 
&\theta_{post}(y_t, y_0)=\tilde{\theta}/\sum_{k=1}^K\tilde{\theta_k},\\
&\tilde{\theta} = [\alpha_ty_t + (1-\alpha_t)/K] \odot [\bar \alpha_{t-1}y_0+(1-\bar \alpha_{t-1})/K],\\
\end{aligned}
\end{equation}
where $\alpha_t = 1-\beta_t$ and $\bar \alpha_t=\prod{s=1}^t\alpha_s$. We utilize $p(y_{t-1}|y_t) = \mathcal{C}(y_{t-1}|\theta_{post}(y_t, \hat{y0}))$ to approximate $q(y_{t-1}|y_t, y_0)$. 
Moreover, the neural network is trained to approximate the noise $\epsilon$ from the noisy input $x_t$ and $\hat{y_0}$ from the noisy sample $y_t$ at timestep $t$. 

Meanwhile, we embrace a non-causal WaveNet architecture as our denoiser and employ a 1x1 convolution layer for the continuous F0 and an embedding layer for the discrete UV.
Finally, We use Gaussian diffusion loss and multinomial diffusion loss to optimize this module.

\subsection{Diffusion Decoder}
\label{sec: appendix1de}

The diffusion decoder uses a 4-step generator-based diffusion model, which parameterizes the denoising model by directly predicting the clean data. 
The 4-step generator-based diffusion model offers the benefits of both excellent perceptual quality and rapid sampling speed.
Meanwhile, the diffusion process is illustrated in Figure \ref{fig: dif}(d).
Like the pitch diffusion predictor, we also use a non-causal WaveNet architecture to be our denoiser.

To train the diffusion decoder, we first apply Mean Absolute Error (MAE) loss.
To be more specific, $x_0$ is the original clean data, while $x_\theta$ denotes the denoised data sample:
\begin{equation}
\begin{aligned}
&\mathcal{L}_{mae} &= \left\|x_\theta\left(\alpha_t x_0 + \sqrt{1 - \alpha_t^2} \epsilon \right) - x_0\right\|,
\end{aligned}
\end{equation}
where $\alpha_t = \prod_{i=1}^t \sqrt{1 - \beta_i}$.
$\beta_t$ represents the predefined fixed noise schedule at diffusion step $t$. 
Additionally, $\epsilon$ is randomly sampled from a normal distribution $\mathcal{N}(0, I)$.

Furthermore, we incorporate the Structural Similarity Index (SSIM) loss as an additional component of the reconstruction loss. 
The SSIM function yields a value between 0 and 1, where a value of 1 indicates the highest similarity to the ground truth, reflecting the best possible performance.
\begin{equation}
\begin{aligned}
&\mathcal{L}_{ssim} = 1 - SSIM\left(x_\theta\left(\alpha_t x_0 + \sqrt{1-\alpha_t^2} \epsilon \right), x_0\right).
\end{aligned}
\end{equation}

\section{Pseudo-Code of the Uncertainty Modeling Layer Normalization}
\label{sec: appendix2}

The algorithm of the UMLN is illustrated in Algorithm \ref{alg: umln}.
\renewcommand{\algorithmicrequire}{\textbf{Input:}}
\renewcommand{\algorithmicensure}{\textbf{Output:}}

\begin{algorithm}[t]
    \caption{Pseudo-Code of the Uncertainty Modeling Layer Normalization}
    \label{alg: umln}
    \begin{algorithmic}
        \REQUIRE{$x$: input content representation of shape (B, T, C), $s$: the addition of the timbre and emotion embedding (B, 1, C), $p$: probability to forward this module, $eps$: a small value added for numerical stability}
        \ENSURE{denormalized input with potential statistics shifts}
        \IF {not in training mode}
        \STATE return $x$
        \ENDIF
        \IF{random probability $>$ p}
        \STATE return $x$    
        \ENDIF
        \STATE Compute the mean and standard deviation of input;
        \STATE $\mu(x)=\frac{1}{C}\sum^C_{c=1}x$
        \STATE $\sigma^2(x)=\frac{1}{C}\sum^C_{c=1}(x-\mu(x)])^2$
        \STATE Normalize input
        \STATE $x_{norm}=\frac{x-\mu(x)}{\sigma(x)+eps}$
        \STATE Get scale and bias
        \STATE $\gamma(s) = E^\gamma*s, \beta(s) = E^\delta*s$
        \STATE Uncertainty estimation 
        \STATE $\Sigma^2_\gamma(s)=\frac{1}{B}\sum^B_{b=1}(\gamma(s)-\mathbb{E}_b[\gamma(s)])^2$
        \STATE $\Sigma^2_\beta(s)=\frac{1}{B}\sum^B_{b=1}(\beta(s)-\mathbb{E}_b[\beta(s)])^2$
        \STATE Compute the synthetic scale and bias randomly sampling from the given Gaussian distributions
        \STATE $\gamma_{um}(s)=\gamma(s)+\epsilon_{\gamma}\Sigma^2_\gamma(s)$
        \STATE $\beta_{um}(s)=\beta(s)+\epsilon_{\beta}\Sigma^2_\beta(s)$
        \STATE Denormalize input using the mixed statistics
        \STATE return $x_{norm}*\gamma_{um}(s)+\beta_{um}(s)$
        
    \end{algorithmic}
\end{algorithm}

\section{Details of Experiments}
\label{sec: appendix3}

\subsection{Subjective Evaluation}
\label{sec: appendix3sub}

We randomly select 16 sentences from the test set for the subjective evaluation. 
Each ground-truth singing voice sample or generated singing voice is carefully listened to by a minimum of 15 esteemed professional listeners. 
For MOS and CMOS evaluations, the listeners are instructed to focus on assessing the audio quality and naturalness while disregarding any differences in styles (such as timbre, emotion, pronunciation, and articulation skills). 
Conversely, for SMOS and CSMOS evaluations, the listeners are instructed to concentrate on evaluating the similarity of styles to the reference, while disregarding differences in content or audio quality. 
In the MOS and SMOS evaluations, each listener is asked to rate different singing voice samples using a Likert scale ranging from 1 to 5. 
In the CMOS and CSMOS evaluations, the listeners are instructed to compare pairs of singing voice samples generated by different systems and indicate their preference, adhering to the following rule: 0 indicates no difference, 1 indicates a slight difference, and 2 indicates a significant difference. 
In the AXY discrimination test, a rater is required to evaluate a reference sample A and two competing samples, X and Y. 
The rater is tasked with assigning a score based on the proximity of X and Y to A. 
The scoring scale ranges from -3 to 3, where a higher score indicates that Y is closer to A compared to X. 
To be specific, -3 to -1 mean “X is much closer", 0 denotes “Both are about the same distance", while 1 to 3 is “Y is much closer".
It is important to note that all listeners receive equal compensation for their participation.

\subsection{Objective Evaluation}
\label{sec: appendix3obj}

We utilize Cosine Similarity and F0 Frame Error (FFE) as objective evaluation metrics to assess the timbre similarity and synthesis quality of the test set. 
Firstly, Cosine Similarity is employed to quantify the timbre resemblance between the synthesized and reference singing voices.
We calculate the average cosine similarity between the embedding extracted from the synthesized voices and the ground truth embedding, providing an objective measure of singer similarity performance. 
Subsequently, FFE combines metrics for voicing decision error and F0 error, capturing crucial F0 information.

\section{Details of Results}
\label{sec: appendix4}

\subsection{Parallel Style Transfer}
\label{sec: appendix4par}

\begin{figure*}[t]
\centering
\includegraphics[width=0.9\textwidth]{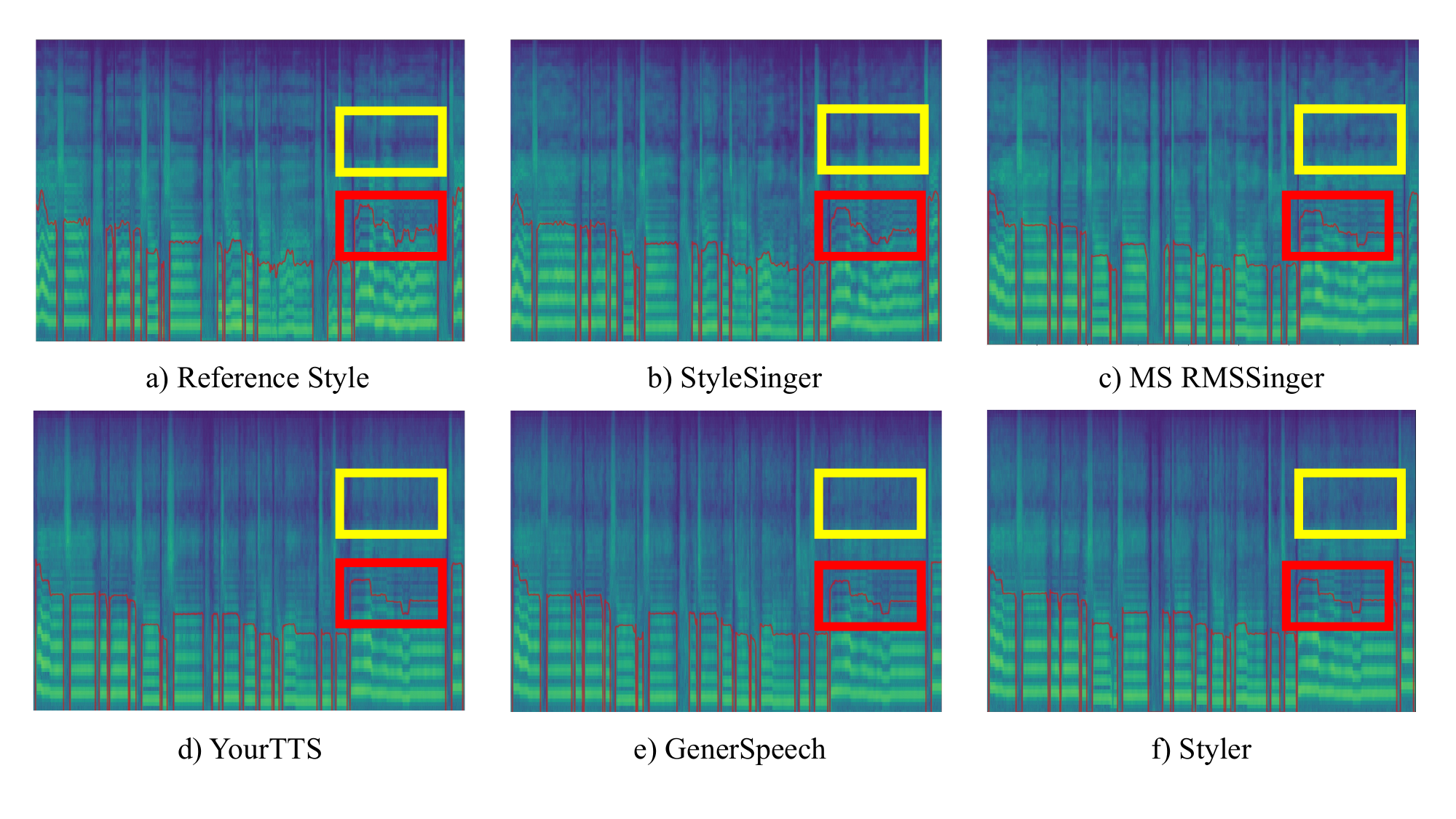}
\caption{The mel-spectrograms are depicting the results of parallel style transfer. 
Red boxes demonstrate that StyleSinger captures the reference style more effectively compared to other baseline models. 
Meanwhile, yellow boxes indicate that StyleSinger produces higher-quality mel spectrograms in the synthesis process.}
\label{fig: par}
\end{figure*}

As shown in Figure \ref{fig: par}, we present the visual results of the parallel style transfer experiment. We observed the following:
1) StyleSinger adeptly captures the stylistic nuances inherent in the reference singing voices. 
The fluctuations and variations in the generated output signify the similarity in vocal techniques.
However, baseline methods demonstrate relatively flat and less expressive curves, indicating a lack of learning of the reference style.
2) StyleSinger demonstrates superior modeling capabilities for mel-spectrograms compared to many other methods, generating high-quality and detailed mel-spectrograms.

\subsection{Ablation Study}
\label{sec: appendix4abl}

\begin{figure*}[t]
\centering
\includegraphics[width=0.9\textwidth]{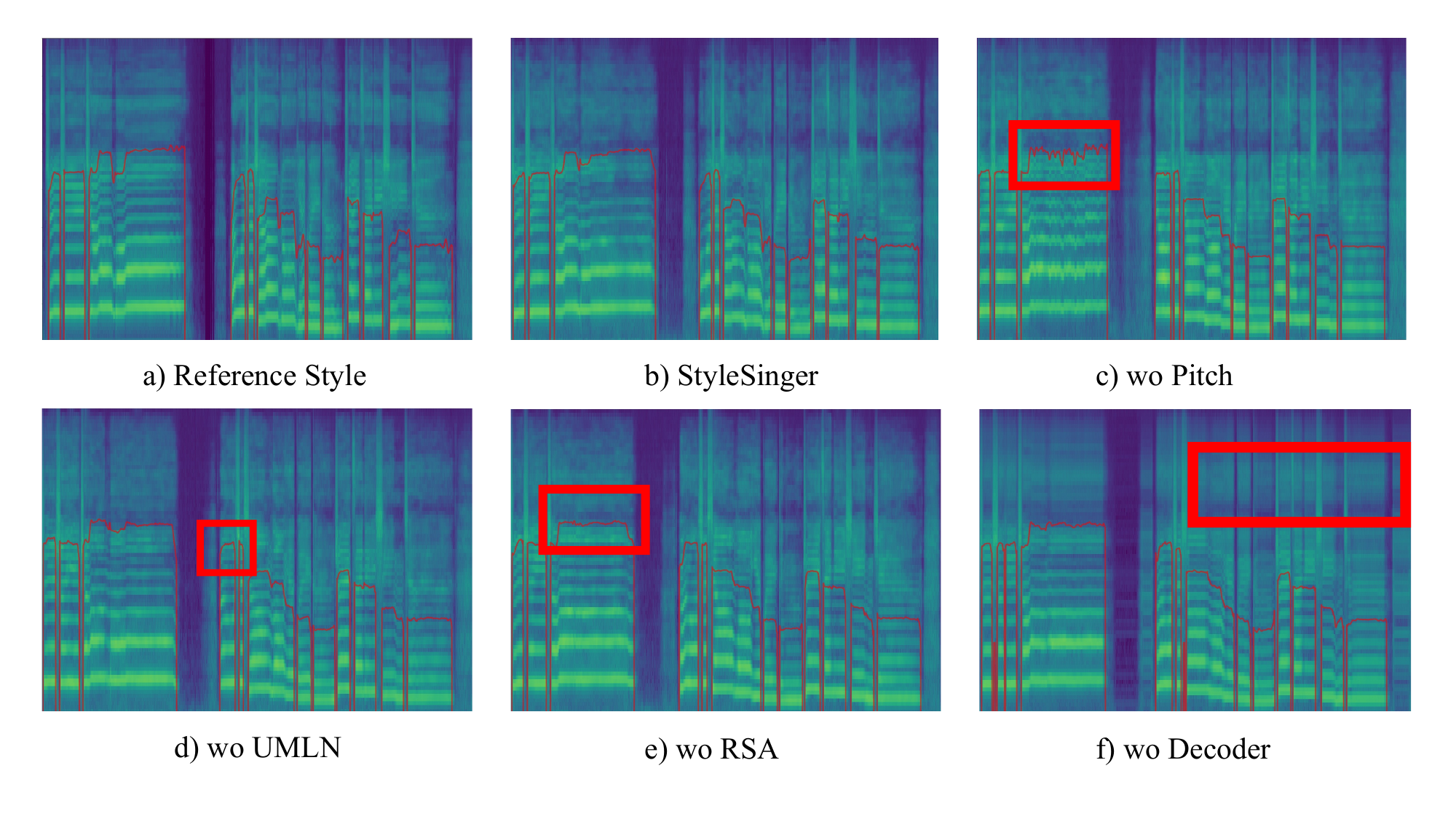}
\caption{The mel-spectrograms depicting the results of the ablation experiment in parallel style transfer.
Red boxes indicate that other models used in the ablation experiments fail to effectively capture the pitch curve or result in a decline in the quality of mel-spectrograms synthesis.}
\label{fig: abl}
\end{figure*}

As shown in Figure \ref{fig: abl}, we present the visual results of the ablation experiment. We observed the following:
1)  Excluding the pitch diffusion predictor, we utilize the simple pitch predictor in FastSpeech2, the model fails to effectively model f0, resulting in drastic fluctuations.
2) When we eliminate the uncertainty modeling layer norm (UMLN), the model's adaptability to out-of-distribution (OOD) scenes deteriorates, resulting in a flat spectrogram curve.
3) As the Residual Style Adaptor (RSA) is removed, the model's ability to capture the styles of the reference samples deteriorates. 
The pitch spectrogram curve lacks the distinctive style fluctuations present in the reference.
4) Without the diffusion decoder, we employ a transformer decoder instead.
The mel-spectrogram becomes unnatural, which denotes that the audio quality generated by the model significantly decreases. 

\end{document}